УДК 004.9 + 520.84

# Идентификация звёзд и цифровая версия каталога Э.С. Бродской и П.Ф. Шайн 1958 года

*М.А. Горбунов[1], А.А. Шляпников[2]*

[1] ФГБУН «Крымская астрофизическая обсерватория РАН», Научный, Крым, Россия, 98409
*mag@craocrimea.ru*
[2] ФГБУН «Крымская астрофизическая обсерватория РАН», Научный, Крым, Россия, 98409
*aas@craocrimea.ru*

**Аннотация.** Рассмотрена процедура идентификации объектов по поисковым картам, определение их координат на эпоху 2000 года и перевод в машиночитаемый формат опубликованной версии каталога Э.С. Бродской и П.Ф. Шайн 1958 года. Приведена статистика фотометрических и спектральных данных оригинального каталога. Описана цифровая версия каталога, его представление в HTML, VOTable и AJS форматах и основные принципы работы с ним в интерактивном приложении Международной виртуальной обсерватории – атласе неба Aladin.

IDENTIFICATION OF STARS AND DIGITAL VERSION OF BRODSKAYA'S AND SHAJN'S CATALOGUE OF 1958, by *M.A. Gorbunov and A.A. Shlyapnikov*. The following topics are considered: the identification of objects on search maps, the determination of their coordinates at the epoch of 2000, and converting into a machine-readable format the published version of Brodskaya's and Shajn's catalogue of 1958. The statistics for photometric and spectral data from the original catalogue is presented. A digital version of the catalogue is described, as well as its presentation in HTML, VOTable and AJS formats and the basic principles of work in the interactive application of International Virtual Observatory – the Aladin Sky Atlas.

**Ключевые слова:** *каталоги, фотометрия, спектры, статистика*

## 1 Введение

Каталог спектров и фотографических звёздных величин 3340 звёзд ярче $12^m.5$ в площадке 45 квадратных градуса созвездия Персея, подготовленный Э.С. Бродской и П.Ф. Шайн (далее В58) был опубликован в 20 томе «Известий Крымской астрофизической обсерватории» (Бродская, 1958). Данная работа выполнялась согласно плану, предложенному академиком Г.А. Шайном, по исследованию звёздной и пылевой составляющей Галактики, изучению пространственного распределения группировок ранних звёзд и выявлению звёзд, возбуждающих галактические туманности (Проник, 2005). Представляя интерес не только в рамках проекта, получившего название «План Шайна», но и собственно как каталог объектов, работа Э.С. Бродской и П.Ф. Шайн, получила ограниченное распространение, как и многие статьи, опубликованные в «Известиях Крымской астрофизической обсерватории» (Шляпников, 2007). В настоящее время (середина 2017 г.) по данным ADS[1], из 3340 звёзд, составляющих описываемый каталог, лишь 9 объектов занесены в базу данных SIMBAD[2]. Последнее обстоятельство послужило основанием для перевода в цифровой машиночитаемый формат каталога Э.С. Бродской и П.Ф. Шайн.

## 2 Структура каталога B58

Статья, в которой напечатан каталог, занимает 45 страниц. В неё входит описание B58 на четырёх страницах, восемь идентификационных карт, и сам каталог с примечаниями на 33-х страницах. Технология перевода B58 в цифровой формат включала в себя сканирование и распознание напечатанной символьной информации специализированным программным обеспечением, аудиовизуальный контроль и редактирование данных.

Каталог состоит из 5-ти столбцов, в которых указаны: порядковый номер звезды в соответствующей зоне склонения; спектральные классы звёзд; видимые фотографические величины $m_{pg}$; показатели цвета $C.I.$, приведенные к системе $B - V$; номера звёзд по каталогу BD (Аргеландер, 1903). Данная структура, с добавлением координат звёзд, была сохранена и для цифровой версии. Примечания, специальные обозначения и ссылки на данные из других каталогов, снабжены соответствующей переадресацией к первоисточнику. Более детальное описание цифровой версии будет дано ниже.

## 3 Идентификации объектов по поисковым картам и определение их координат

Основной проблемой при создании современного интероперабельного варианта B58 стало отсутствие координат каталогизированных объектов. Каталог разделен на шесть зон по склонению, внутри каждой из которых звезде присвоен соответствующий зоне номер. Данное обозначение с указателем есть на идентификационной карте, по которой, в классическом случае, необходимо найти объект для последующего извлечения информации о нём из B58. Очевидно, что подобная практика, широко распространённая в середине прошлого века, на сегодняшний день неприемлема.

Идентификация и определение координат звёзд выполнялось по методике, предложенной А.А. Шляпниковым с помощью интерактивного атласа неба Aladin (Боннарель, 2000), предназначенного для визуализации оцифрованных астрономических изображений и послойного наложения на них информации из астрономических каталогов и баз данных. Рисунки с отождествлением звёзд из каталога B58, были отсканированы, загружены в Aladin и для них выполнена астрометрическая калибровка. Последнее позволило совместить с идентификационными картами B58 информацию о положениях звёзд из других каталогов. Рисунок 1 показывает фрагмент зоны +56° в области прямого восхождения $02^h29^m$. Здесь, на части иллюстрации 2 из статьи Е.С. Бродской 1958 года, маркерами отмечены положения звёзд из каталога Tycho-2 (Хог, 2000). Если в Tycho-2 отсутствовала звезда из B58, координаты для неё брались из каталога APASS (Хенден, 2016). Для создания файла с координатами звёзд необходимо было открыть редактор в консоли Aladin и поочерёдно, кликая мышкой по маркеру у звезды соответствующего номера, добавлять его к появляющимся в окне редактора координатам.

---

[1] B-1958 в ADS // http://adsabs.harvard.edu/abs/1958IzKry..20..299B

[2] B-1958 в SIMBAD // http://cdsbib.u-strasbg.fr/cgi-bin/cdsbib? 1958IzKry..20..299B

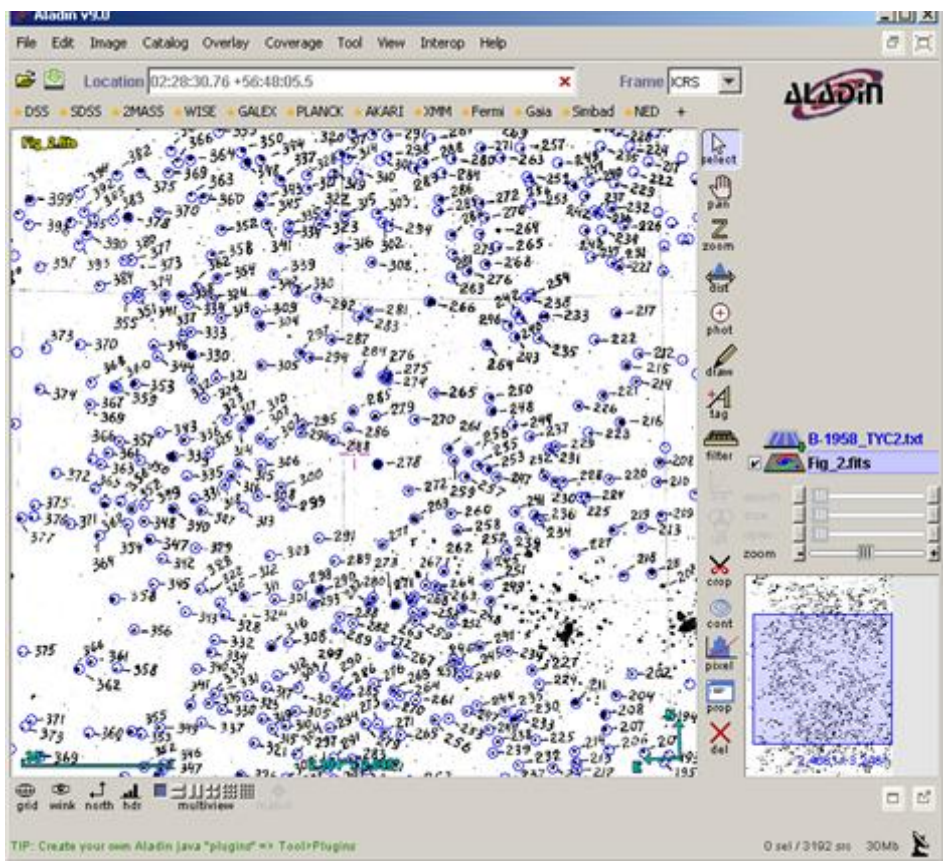

**Рис. 1.** К определению координат звёзд из B58 (пояснения в тексте).

Для каждой зоны B58 был создан отдельный текстовый файл с координатами и номерами звёзд. К нему, по соответствующим номерам, путём перекрёстной идентификации, добавлялась информация из оригинального каталога. После создания сводного по зонам каталога B58 он был проверен на повторяющиеся координаты у звёзд с разными номерами. После учёта дублирующих координат (ошибок в их определении) был полученный файл – прототип цифровой версии каталога Э.С. Бродской и П.Ф. Шайн. Преобразование его формат, поддерживаемый Aladin позволило провести более детальный анализ объектов, вошедших в B58.

## 4 Статистика фотометрических и спектральных данных

Прежде, чем приступать к сравнению информации из каталога B58 с другими каталогами мы провели статистический анализ вошедших в него данных. При этом не были исключены данные, заимствованные авторами из других каталогов, а также неуверенные определения.

Рисунок 2 иллюстрирует распределение объектов каталога B58 по звёздным величинам в полосах $B$ (левая панель) и $V$ (правая панель). Лишь для 2115 звёзд из 3400 каталога приведены показатели цвета $B - V$, на основании которых, для последующего анализа, рассчитывалась величина $V$. В описании каталога приведена его внутренняя точность, которая характеризуется среднеквадратической ошибкой в определении $B$ и $V$ величин, как $\pm 0^m,034$ и $\pm 0^m,036$, соответственно. Этот диапазон ошибок был использован нами при дальнейшем анализе звёзд на возможную переменность, в сравнении с данными из других каталогов.

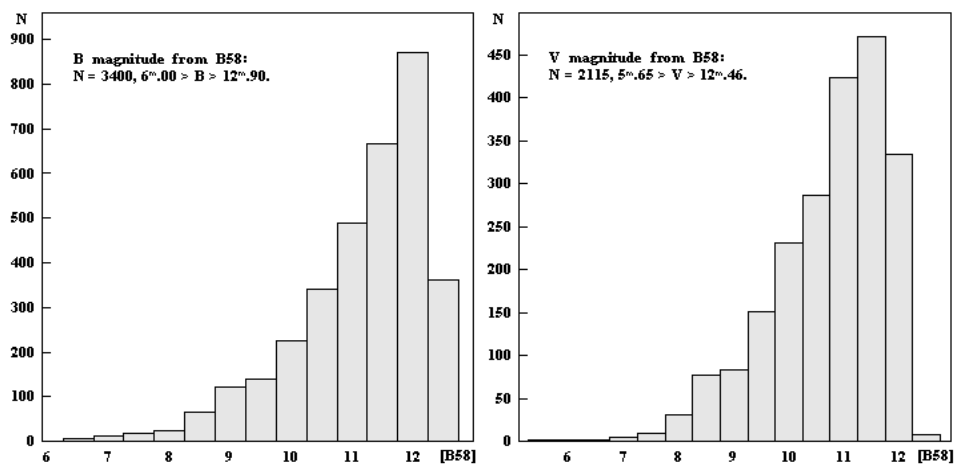

**Рис. 2.** Распределение объектов B58 по звёздным величинам.

Максимум в распределении ***B*** величин приходится на 12 звёздную величину, а для ***V*** величин он смещён в область более ярких звёзд. Очевидно, что прямые снимки, по которым находились ***B*** и ***V*** величины объектов имели большую проницающую способность. Однако определяющим фактором в данном случае служило то, на сколько хорошо был виден и классифицируем спектр на негативах, полученных с объективной призмой.

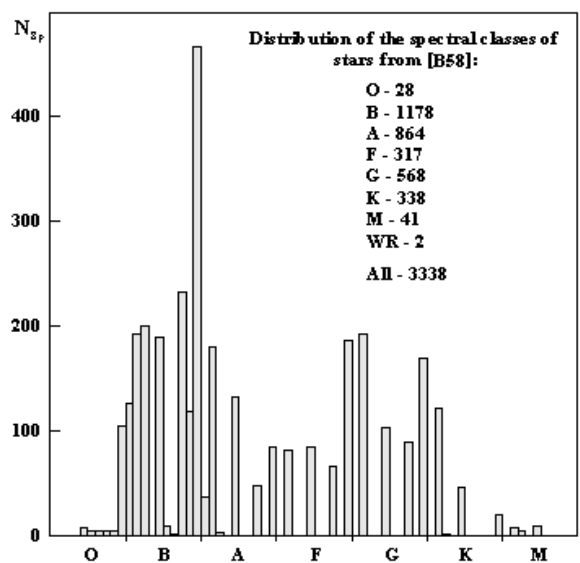

**Рис. 3.** Распределение объектов B58 по спектральным классам.

На рисунке 3 представлено распределение числа звёзд в зависимости от приведённых в каталоге B58 спектральных классов и подклассов. Наибольшее число звёзд – 2042 имеют спектральную классификацию **B** и **A**. Два объекта в каталоге классифицированы, как звёзды Вольфа-Райе, а для звёзд № 14 и № 184 в зоне +57° спектральная классификация отсутствует.

## 5 Цифровая версия каталога B58

Принципы работы с сетевыми версиями каталогов, подготовленных в КрАО, описаны ранее (Шляпников, 2013). Цифровая версия B58 представлена (рис. 4) в базе данных каталогов и списков проекта «План Шайна» на сайте Крымской астрономической виртуальной обсерватории (КрАВО). В структуре КрАВО, для описания и работы с каталогами предусмотрен ряд закладок на странице, содержащей информацию о них. В закладке *«Abstract»* - указаны название, автор(ы), ключевые слова, год, том и первая страница публикации в «Известиях Крымской астрофизической обсерватории», комментарии о числе библиографических ссылок, таблиц и рисунков, библиографический код публикации по классификатору SAO/NASA ADS с гиперссылкой к этой базе данных, гиперссылка на полный текст публикации в PDF и GIF форматах. Закладки *«Publication»*, *«Identification Map»*, *«Catalogue»* содержат, соответственно, отсканированную версию публикации, идентификационной карты, если она есть, и каталога. В закладке *«References»* приведен список литературы из публикации с гиперссылками на базу данных SAO/NASA ADS. *«Digital version»* - закладка, представляющая собой набор ссылок к информации собственно цифровой версии каталога. В закладке *«Analysis»* содержатся результаты анализа каталога B58 в сравнении с современными данными. Этому разделу посвящена отдельная статья, опубликованная в данном томе.

**Рис. 4.** Представление закладки «Digital version» каталога B58.

## 6 Описание закладки «Digital version» каталога B58

Закладка разделена на три колонки. В каждой из них содержатся гиперссылки к HTML файлам с соответствующей информацией цифровой версии каталога. Первая колонка обеспечивает доступ к общему описанию B58. Ссылка *«Description»* даёт описание каталога по колонкам. В первой и второй колонках приведены координаты звёзд определённые по методике, описанной в разделе **3**. Далее в колонках информация, описанная в разделе **2**. Последняя колонка, как и первые две, является дополнением к оригинальной версии каталога. В ней приведены комментарии о конкретных объектов, представленные в «Примечании» оригинального каталога, а также указаны объекты, выделенные в B58 жирным шрифтом с указанием ссылки на публикацию, из которой была заимствована информация. Ссылка *«Summary»* содержит информацию об обозначении B58 в базе данных КрАВО, названии, авторах, публикации, библиографическом коде, ключевых словах, статистику о числе объектов, размерах в килобайтах HTML, XML и DATA файлов. Ссылка *«ReadMe»* соответствует формату, применяемому в базе данных VizieR для описания, представленных в ней каталогов. На данной странице: идентификатор каталога по VizieR, его название, авторы, библиографический код по SAO/NASA ADS, ключевые слова, абстракт, описание, побайтовое описание структуры каталога, комментарии, благодарности, библиографические ссылки.

Ссылка *«Identification Map»* в закладке *«Digital version»* содержит таблицу, в которой идентификационные карты каталога B58 представлены в двух форматах *fits* и *ajs*. В формате *fits* изображения имеют астрометрическую калибровку и были использованы для определения координат звёзд (см. раздел **3**). Формат *ajs* (Aladin Java Script) позволяет загружать идентификационную карту в интерактивный атлас неба Aladin с указанием звёзд из каталога B58. Данные на странице *«Identification Map»* представлены в виде таблицы с указанием границ изображений по прямому восхождению и склонению – соответственно. Их загрузка в Aladin производится путём сохранения файлов на персональном компьютере, с последующим открытием в интерактивном атласе, либо копированием сетевого пути (гиперссылки) в окно «Location» в Aladin.

*«Digital Plates Archive»* - раздел, обеспечивающий доступ к оцифрованному архиву негативов стеклянной библиотеки КрАО для данного каталога, в котором изображения представлены в трёх форматах: *jpg*, *ajs* и *fits*. Ссылка *«JPG»* позволяет перейти к странице цифровой версии каталога B58 содержащей уменьшенные копии оригинальных негативов, полученных по «Плану Шайна». Файлы *ajs* и *fits*, загружаемые по соответствующим ссылкам, позволяют отображать в Aladin область покрытия неба пластинками и сами отсканированные негативы (рис. 5).

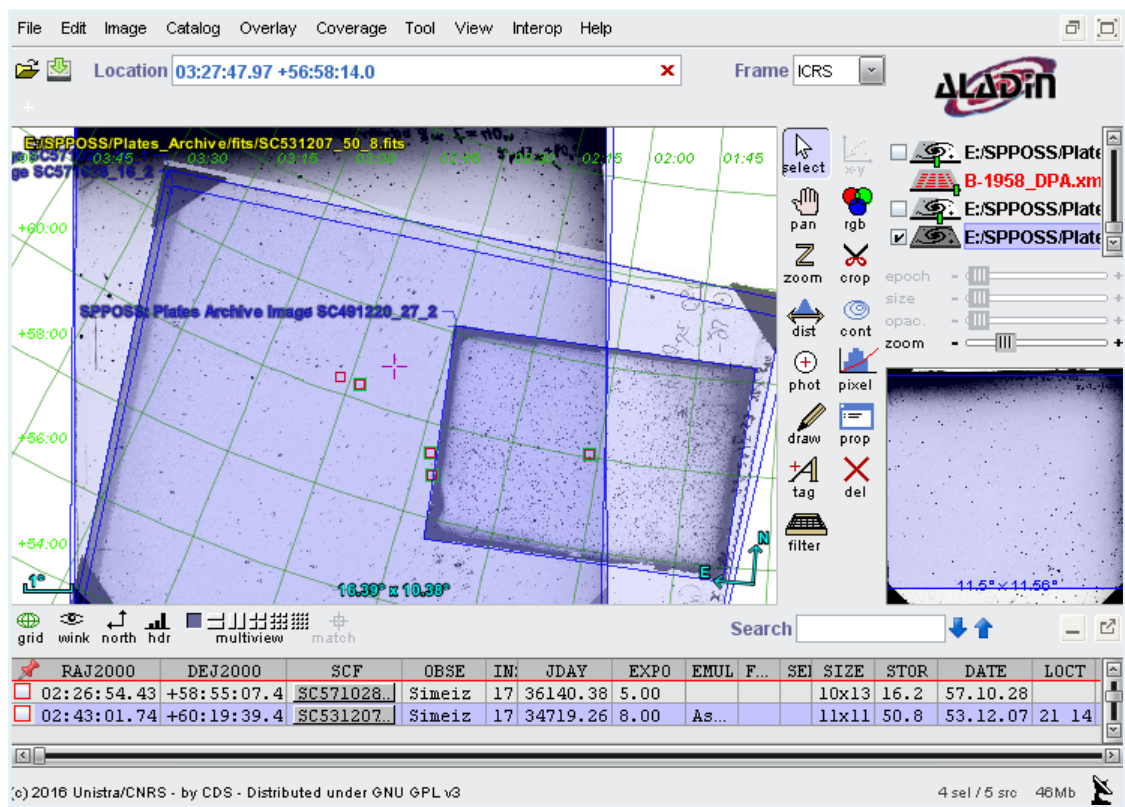

**Рис. 5.** Представление загрузок файлов *ajs* и *fits* в Aladin.

Центральная часть закладки *«Digital version»* демонстрирует распределение объектов B58 на небе и обеспечивает поиск объектов по их идентификаторам или координатам с помощью интерактивного атласа Aladin. Специальный блок, написанный на языке JavaScript, и интегрированный в HTML код, обеспечивающий интерактивную работу с B58, производит переадресацию и открывает в новой вкладке браузера апплет Aladin с загрузкой XML версии каталога (см. ниже) центрированного на введённый запрос.

Правая колонка закладки *«Digital version»* содержит три раздела с доступом к цифровым версиям каталога B58 в различных форматах. В разделе *«Full catalogue»* представлена HTML версия, которую можно загрузить в текущем *«cw»* или в новом окне *«nw»* браузера. Гиперссылка *«VOTABLE»* – версия B58 в формате, разработанном для приложений Международной виртуальной обсерватории (включает XML структуру). Возможно сохранение каталога в данном формате на персональном компьютере, или сетевое открытие его в приложениях виртуальной обсерватории (например, Aladin). DATA версия представляет собой ASCII форматированный файл с данными каталога B58, который открывается в новом окне (ссылка *«nw»*). Здесь же загружается файл, содержащий комментарии (ссылка *«Note»*) к цифровой версии каталога.

Раздел *«Full catalogue with»* обеспечивает использование цифровой версии B58 с апплетом Aladin (ссылка *«ALADIN»*, которая открывается в новом окне браузера), либо копированием гиперссылки (строка ниже) в окно «Location» в программу Aladin, установленную на компьютере пользователя. *«SIMBAD»* и *«VizieR»* – переадресация к соответствующим базам данных. В первом случае – к уже интегрированной информации из B58. Во втором случае – после добавления в базу данных VizieR.

В разделе *«Full catalogue publication»* обеспечены ссылки к цифровой версии каталога B58 в GIF и PDF форматах с указанием размеров соответствующих файлов.

## 7 Заключение

Рассмотренная в статье процедура идентификации объектов B58 по поисковым картам и определение их координат на эпоху 2000 года будет применена при создании цифровых версий других каталогов, полученных по «Плану Шайна». Также будут сохранены структуры HTML, VOTable и AJS форматов, что обеспечит единообразие представления данных.

## Литература